\documentclass{PoS}

\usepackage{feynmp}

\title{A novel method for evaluating correlation functions in lattice hadron spectroscopy\footnote{Based on talks given by Justin Foley and Chik Him Wong.}}

\ShortTitle{}

\author{{Justin Foley\footnote{Present address: Dept. of Physics at the University of Utah, and The Center for Computational Sciences, University of Tsukuba.}, Chik Him Wong}  \\ 
	Dept. of Physics, Carnegie Mellon University, Pittsburgh, PA 15213, USA }
	
\author{John Bulava \\
	NIC, DESY, Platanenallee 6, 15738, Zeuthen, Germany }

\author{K. Jimmy Juge \\
	Dept. of Physics, University of the Pacific, Stockton, CA 95211, USA}

\author{David Lenkner, Colin Morningstar \\ 
	Dept. of Physics, Carnegie Mellon University, Pittsburgh, PA 15213, USA}

\author{Michael Peardon \\
	School of Mathematics, Trinity College, Dublin 2, Ireland}

\abstract{We describe a new approach for evaluating hadronic correlation functions
	which combines Laplacian-Heaviside quark smearing with a stochastic estimator of quark propagators.
        This method utilizes noise dilution in a new way to reduce the variance in correlators.
	The efficacy of the new algorithm is demonstrated on a number of systems, including disconnected 
	diagrams and multi-hadron correlators, on a small lattice where comparisons with the results obtained with exactly determined quark propagators are possible. On larger lattice volumes, the use of exact propagators becomes prohibitively expensive, while the stochastic
	method is still computationally feasible.}

\FullConference{The XXVIII International Symposium on Lattice Field Theory, Lattice2010\\
		June 14-19, 2010\\
		Villasimius, Italy}

\begin{document}

\section{Introduction}
A long-term aim of our collaboration is the first-principles determination
of stable hadron masses, and resonance energies and decay widths. 
To date, considerable progress has been made in extracting finite-volume stationary-state 
energies in the isovector meson, kaon, and baryon sectors~\cite{nf2nuc,nf2nuc2,dudek}. 
These initial studies, which were performed at relatively heavy pion masses and moderate lattice volumes, 
involved interpolating operators specifically designed to couple strongly to 
single-hadron states.  
However,
at lighter pion masses and larger spatial volumes,
explicit multi-hadron operators must be incorporated into the analysis.
Until recently, the inclusion of multi-hadron interpolators has been problematic 
because the resulting two-point functions may involve quark lines which begin and 
end on the sink time slice and source operators on all spatial sites of a time slice.
Hence, a general treatment of multi-hadrons is not 
amenable to conventional point-to-all quark-propagator techniques.
In analyses involving single-particle operators, 
disconnected contributions to flavor-singlet 
meson correlators are similarly problematic, and 
little progress has been made to date in determining the $I=0$ 
meson spectrum from Monte Carlo simulations.

In this article, we describe a novel method~\cite{hadron2009} for evaluating hadronic 
correlation functions, which combines quark-field smearing with a stochastic estimator. 
The new method facilitates the precise evaluation of 
multi-hadron and flavor-singlet meson correlators, at a significantly lower computational 
cost than previous approaches. 
First results obtained on intermediate lattice volumes are presented, and we briefly 
discuss more recent studies performed on  larger lattices.

\section{An alternate quark-field smearing scheme}
\begin{figure}
\begin{minipage}[b]{0.5\linewidth}
\centering
\includegraphics[scale=0.38]{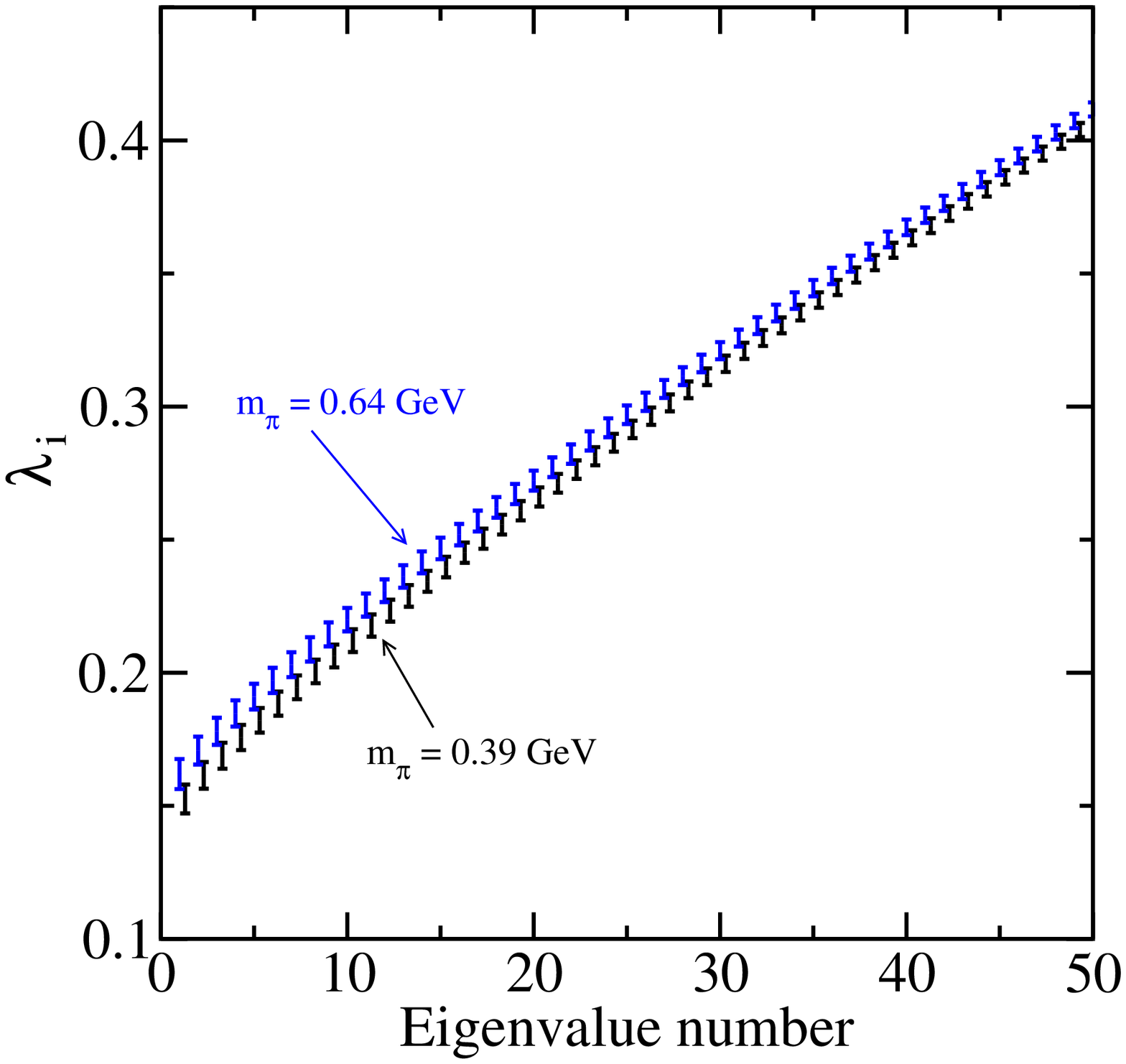}
\end{minipage}
\hspace{0.1cm}
\begin{minipage}[b]{0.5\linewidth}
\centering 
\includegraphics[scale=0.38]{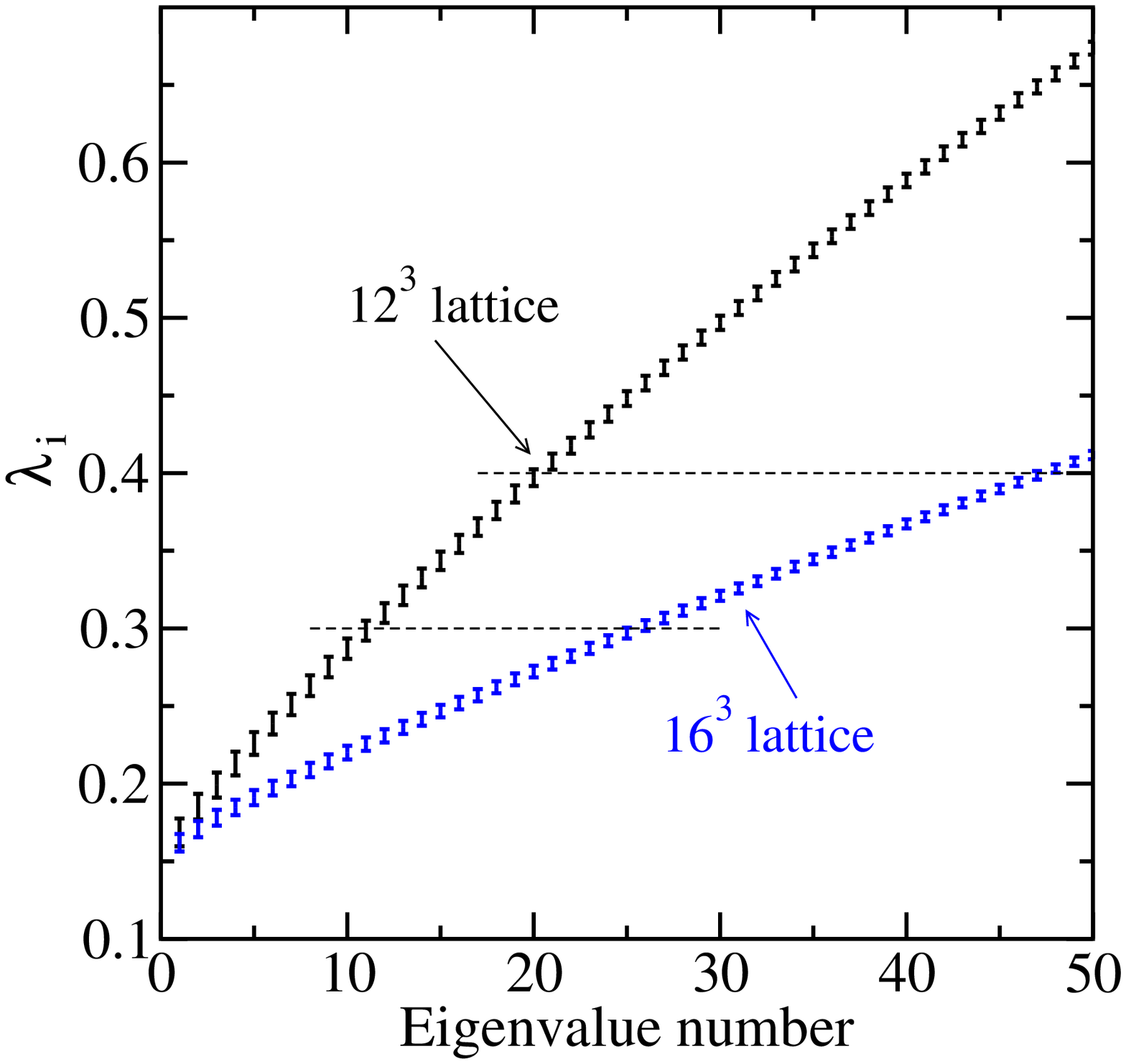}
\end{minipage}
\caption{
	The plot on the left shows the Monte Carlo estimate of low-lying $\lambda_i$'s
	computed on a $16^{3}$ spatial lattice on dynamical ensembles 
	with two different pion masses using stout-smeared link variables. 
	The pion-mass dependence of the Laplacian spectrum is seen to 
	be very mild. The right-hand plot shows Laplacian spectra 
	computed on 2+1 flavors simulations performed on two different 
	lattice volumes. All other simulation parameters match. 
	The dashed horizontal lines are guides for the eye, showing 
	that the density of eigenmodes grows linearly with the 
	spatial lattice volume.
      }
\label{lambda}
\end{figure}

Our approach relies on the fact that generally in a spectroscopy calculation,  
gauge-covariant smearing is applied to the quark fields 
in  hadron interpolators in order to reduce the coupling of 
these operators to very high-lying states.
In the resulting correlation functions, 
the quark propagator $M^{-1}$ is always sandwiched between 
smearing operators: $S M^{-1} S $, where $\tilde{\psi} = S \psi $ is 
the smeared quark field.
We define Laplacian-Heaviside~\cite{distillation}, or LapH, quark-field smearing by
\begin{eqnarray}
S = \Theta \left( \tilde{\triangle} + \sigma^{2} \right) 
,\hspace{5mm} S_{ab}(x,y) \approx \delta_{x_4,y_4}\sum_{k=1}^{N_v} v_a^{(k)}\left( x \right) v_b^{(k)}\left( y \right)^*, 
\label{eqn:LapH}
\end{eqnarray}
with 
\begin{eqnarray}
	\tilde{\triangle}_{ab}(x, y ) v^{(i)}_b \left( y \right) = -\lambda_i(x_4) v^{(i)}_a(x),
\hspace{5mm} v^{(i) *}(x)v^{(j)}(x) = \delta^{i j} ,
\end{eqnarray}
where $\tilde{\triangle}$ is a gauge-covariant Laplacian operator constructed from 
stout-smeared~\cite{stoutcolin} link variables $\tilde{U}$ :
\begin{eqnarray}
	\tilde{\triangle}_{ab}(x,y)=\sum_{k=1}^3 \left\{ \tilde{U}_k^{ab}(x)\delta(x+\hat{k},y)+\tilde{U}_k^{\dagger ab}(y)\delta(x-\hat{k},y)-2\delta(x,y)\delta^{ab}\right\}. 
\end{eqnarray}
The use of smeared link variables in the quark-field smearing operator has been shown to significantly 
reduce the variance in hadronic correlation functions~\cite{stout}.
The eigenvalues of the Laplacian operator, $-\lambda_{i}$, are negative(see Fig.\ref{lambda}), and the Heaviside function in Eq.~\ref{eqn:LapH} 
imposes a cutoff ( parametrized by $\sigma^{2}$ ) on high-momentum modes.
The level of quark smearing is increased by lowering the cutoff, thereby excluding more Laplacian eigenmodes. 
To evaluate hadronic correlation functions involving LapH-smeared 
quark fields, one does not need to compute the quark-propagator components
directly, instead only the matrix elements of the propagator 
between a subset of Laplacian eigenvectors are required.
For a high enough level of quark smearing, i.e., a small enough subset of 
eigenvectors, it is feasible to compute 
hadron correlation functions which are currently beyond the reach of 
conventional lattice techniques~\cite{distillation}.
In practice, however, the volume dependence of the Laplacian eigenmode distribution is 
a significant limitation when computing $SM^{-1}S$ exactly. 
Although the quark-mass dependence of the Laplacian spectrum is mild, 
the density of eigenmodes scales linearly with the spatial lattice volume.     
Therefore, while an exact treatment of $SM^{-1}S$ may work well on smaller 
lattice volumes as it stands, the increased density of eigenmodes
makes this approach impractical on larger lattices.

\subsection{Stochastic estimation}
To mitigate the volume dependence, the LapH smearing scheme can be 
combined with a stochastic estimator. To proceed, we write
\begin{eqnarray}
S M^{-1} S = S M^{-1} |v \rangle E \left( \eta \eta^{\dagger}  \right) \langle v |.
\end{eqnarray}
$E \left( \eta \eta^{\dagger} \right)$ denotes an expectation value over the outer product of 
noise vectors whose components satisfy 
\begin{eqnarray}
E \left( \eta_{i \alpha} (t) \right) = 0,
  \hspace{5mm}
E ( \eta_{i \alpha} \left(t \right) 
    \eta^{*}_{j \beta} \left( t' \right ) ) = \delta_{i j} \delta_{\alpha \beta} \delta_{t t'}, 
\end{eqnarray}
where $i (j)$ and $\alpha (\beta)$ are eigenmode and spin indices,  
respectively. Note that the noise vectors have neither color nor spatial 
indices.
The number of quark-matrix inversions required to estimate the 
smeared quark line in this way is determined by the number 
of stochastic vectors employed rather than the number of Laplacian 
eigenvectors used in the smearing. However, in general, this naive estimate 
for the smeared quark line is too noisy to be of practical use, and 
we must apply a variance reduction scheme to obtain an improved stochastic 
estimate.

In a number of systems, noise partitioning, or dilution, 
turns out to be a particularly effective 
variance reduction technique~\cite{dilution}. 
Dilution is implemented by choosing noise vectors whose components 
have unit norm, i.e., $Z_{n}$ or $U(1)$ noise, 
and partitioning the noise vector indices, eigenmode, time and spin, 
into disjoint sets. 
To each set of indices, $d$, we assign a projection operator 
$P^{[d]}$ which acts on the noise vectors, with 
components 
\begin{eqnarray}
P^{[d]}_{i j \alpha \beta} \left(t, t' \right) &=& 1  \hspace{2.5mm}  \textrm{if $\left(t,i,\alpha \right)$ and $\left(t,j,\beta \right) \in d$}, \nonumber \\
P^{[d]}_{i j \alpha \beta} \left( t,t' \right) &=& 0  \hspace{2.5mm} \textrm{otherwise.}
\end{eqnarray}
The naive noise average $E \left( \eta \eta^{\dagger} \right)$ is 
replaced by an expectation value involving diluted noise 
vectors $ \sum_{d} E \left( \eta^{[d]} \eta^{[d] \dagger} \right) $, 
 where $\eta^{[d]} = P^{[d]} \eta$.
In the maximal dilution limit, the exact smearing scheme is recovered.
We refer to the combination 
of the LapH smearing scheme with a dilute stochastic estimator as the 
Stochastic LapH method.

\section{Implementation}
In practice, the estimates for the smeared quark propagator are 
implemented in terms of pseudofermion fields which we call 
quark-line ends. For each diluted noise vector $\eta^{[d]}$, we 
form a quark-line source $\rho^{[d]}$, with components   

\begin{eqnarray}
\rho^{[d]}_{\alpha a} \left( x \right)
= \sum_{i}  \eta^{[d]}_{i \alpha} \left(x_4 \right) v_{i a} \left( x \right).
\end{eqnarray}
For each source field, there is a corresponding quark-line sink
$ \phi^{[d]}  = S M^{-1} \rho^{[d]}. $
The smeared quark line can then be written
\begin{eqnarray}
S M^{-1} S = \sum_{d} E \left( \phi^{[d]}  \rho^{[d] \dagger} \right).
\label{eq:quarkline1}
\end{eqnarray}
In addition, the fact that the quark propagator satisfies
\begin{eqnarray}
M^{-1} \left( y; x \right) = \gamma_{5} \left[ M^{-1} \left( x; y \right) \right]^{\dagger} 
\gamma_{5}
\end{eqnarray}
and the hermiticity of the quark smearing operator lead to a 
second expression for the smeared quark line
\begin{eqnarray}
S M^{-1} S = \sum_{d} E \left( \overline{\rho}^{[d]} \overline{\phi}^{[d] \dagger} \right),
\label{eq:quarkline2}
\end{eqnarray}
where $ \overline{\rho}^{[d]} = \gamma_{5} \rho^{[d]} $ and 
$ {\overline{\phi}}^{[d]} = \gamma_{5} \phi^{[d]} $.

On a typical lattice, the complete set of pseudofermion 
fields needed to implement the quark-line estimate is too large to 
save to disk. Instead, to store the quark-line estimates on a 
given configuration, we save the set of low-lying Laplacian 
eigenvectors used in the smearing, the diluted noise vectors $\eta^{[d]}$
\footnote{In practice, we simply record a random initial seed, the dilution scheme, and a rule for generating the noise vector from the seed.} , 
and, for each noise vector, a tensor with elements
\begin{eqnarray}
\Pi \left( i, \alpha, x_4; d \right) = 
\sum_{\mathbf{x}~a} v_{i a}^{*} \left(x \right)   \phi^{[d]}_{\alpha a} (x).
\end{eqnarray}
Using these components, the pseudofermion source and sink fields can be quickly reconstructed when needed. 
Only the Laplacian eigenvectors require a significant amount of storage, which is however independent of the 
number of quark lines to be estimated and the number of noise vectors and the dilution schemes 
used.

To illustrate how the quark-line ends can be combined to form a 
hadron correlator, consider the simple meson two-point function
\begin{eqnarray}
	C^{mn} \left( t \right) 
= \langle 0 | \tilde{\overline{\psi}} \Omega_m \tilde{\psi} \left( t\right) 
 \tilde{\overline{\psi}}  \Omega_n^\dagger  \tilde{\psi} \left(0 \right) |0 \rangle,
\end{eqnarray}
where $\Omega_m$($\Omega_n$) is an arbitrary combination of gauge-covariant spatial 
displacement operators and spin matrices acting on the quark fields.
For compactness, all indices apart from the time label have 
been suppressed. 
Wick contracting the quark fields, and using both 
Eq.~\ref{eq:quarkline1} and Eq.~\ref{eq:quarkline2}, 
we obtain an estimate for the connected component 
of this correlator
\begin{eqnarray}
	C_{\textrm{\tiny{conn}}}^{mn} \left( t \right)
&\approx& - 
\Big \langle
\frac{1} {N_{r} N_{s}}\sum_{r \neq s}
\sum_{d_{r}~d_{s}}  
\left [ 
\overline{\phi}^{[d_{r}] \dagger}_{r} \Omega_m \phi^{[d_{s}]}_{s} \left( t \right)  \rho^{[d_{s}] \dagger}_{s} \Omega_n^\dagger \overline{\rho}^{[d_{r}]}_{r} \left( 0 \right)    
\right]
\Big \rangle, 
\label{eq:correlator}
\end{eqnarray}
where the superscripts $r$ and $s$ label different noise sources and $\langle ... \rangle$ denotes an average over the gauge ensemble.
Other estimates for the connected correlator, involving 
different combinations of `barred' and `unbarred' quark line ends,
are also possible.
However, since Eq.~\ref{eq:correlator} involves the quark sources 
$\rho_{r}^{[d_{r}]}$, $\rho_{s}^{[d_{s}]}$ on a single time slice, $t=0$, 
only, it has the 
advantage that both quark-line estimates are automatically 
fully diluted in time.

Defining the mesonic `operators' 
$\mathcal{O}_{\Omega_m [ \bar{r}~s]}^{[d_{r}~d_{s}]}  \equiv 
\overline{\phi}_{r}^{[d_{r}] \dagger} \Omega_m \phi_{s}^{[d_{s}] } $ 
and $\overline{\mathcal{O}}_{\Omega_n [ \bar{r}~s]}^{[d_{r}~d_{s}]}  \equiv \left(\overline{\rho}_{r}^{[d_{r}] \dagger} \Omega_n \rho_{s}^{[d_{s}]}\right)^\dagger = \rho_{s}^{[d_{s}] \dagger} \Omega_n^\dagger \overline{\rho}_{r}^{[d_{r}]} $ 
allows the connected correlator to be written in a more compact 
form
\begin{eqnarray}
	C_{\textrm{\tiny{conn}}}^{mn} \left( t \right) 
\approx 
-
\Big \langle 
\frac{1} {N_{r} N_{s}}
\sum_{r \neq s} \sum_{d_{r}~d_{s}} 
\mathcal{O}_{\Omega_m [ \bar{r}~s] }^{[d_{r}~d_{s}]} \left( t \right)
\overline{\mathcal{O}}_{\Omega_n [ \bar{r}~s] }^{[d_{r}~d_{s}]} \left( 0 \right)
\Big \rangle.
\end{eqnarray}
The factorization of the correlator estimates into contributions 
from different hadron operators is an extremely useful feature of the method.
It facilitates the construction of correlator 
matrices involving large sets of interpolating operators and simplifies the 
evaluation of multi-hadron correlators.


\section{First tests}
\begin{figure}
\centering
\scalebox{0.38}{\includegraphics*{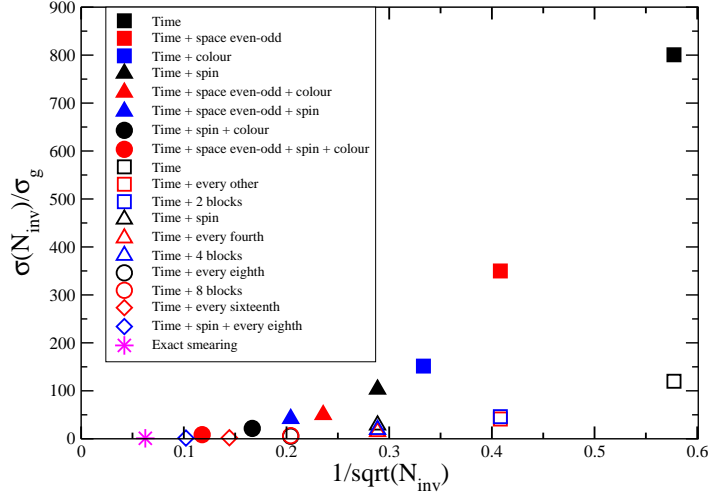}}
\caption{Ratio of standard deviation on time slice 5 of a triply-displaced nucleon correlator
evaluated using different stochastic estimators to the standard deviation on the 
exact LapH estimate of the correlator at the same time separation. The filled symbols denote
measurements obtained using noise introduced on the entire lattice and the open symbols are results from the Stochastic LapH approach.
For every dilution scheme considered, Stochastic LapH significantly outperforms 
the lattice noise approach \cite{jbthesis}.}

\label{dilutioncomparison}
\end{figure}

Tests of the Stochastic LapH algorithm on connected hadron correlators,
such as isovector-meson and baryon two-point functions, have shown that, 
with a judiciously chosen dilution scheme, 
this method yields statistical errors that are comparable to the errors 
obtained using an exact treatment of $SM^{-1}S$ , at a considerably lower computational cost.
Moreover, the new method is more 
efficient than using diluted stochastic estimates with noises introduced on the entire lattice, as described in Ref.~\cite{dilution}. 
This  is  clearly demonstrated in Fig.~\ref{dilutioncomparison}, 
which shows the ratios of standard deviation on a single time slice of
 a nucleon correlator evaluated 
using different stochastic methods 
to that obtained from exact LapH. 
The solid symbols denote the results obtained by introducing noise on the lattice. The open symbols 
are results from the Stochastic LapH method. The nucleon operator 
in question has gauge-covariant displacements in three directions.
These measurements were performed on $100$ $2+1$ flavor gauge-field configurations
with a pion mass of approximately 400~MeV, on a $20^{3} \times 128$ anisotropic lattice. The spatial lattice spacing was approximately $0.12$~fm,
and the ratio of spatial to temporal lattice spacings was $3.5$.
The smearing uses the $64$ lowest-lying Laplacian eigenmodes on each time slice.
The statistical uncertainties on the correlators were estimated using the 
Jackknife algorithm. The data here correspond to a temporal 
separation between the source and sink operators of $t = 5 a_{t}$,  
although the same qualitative behavior is observed for other temporal separations.
The minimal level of dilution included in this figure is 
full time dilution. The reason is that for a connected correlator evaluated on a single source time slice, the inversion cost is the same regardless the time dilution scheme. Any non-full time dilution schemes would only impose larger variance from the noises without any reduction of computational costs.
The eigenmode dilution schemes considered include partitioning the 
Laplacian eigenmodes into blocks of adjacent modes, as well as choosing 
subsets consisting of eigenmodes that are separated by some number $n$ 
of intermediate modes, known as interlacing.
The Stochastic LapH estimates use three independent $Z_{4}$ noise vectors - 
the minimum number required to obtain unbiased stochastic estimates. 
It is possible to improve the correlator estimates by averaging over different 
orderings of the noise vectors. However, this average has not been performed 
on the examples presented here.


For a given number of quark-matrix inversions, the error on the correlator 
computed using Stochastic LapH is much smaller than the error on the 
correlator estimate involving stochastic quark propagators. 
Fig.~\ref{dilutioncomparison} also shows that there is no significant difference between using blocked eigenmode dilution 
projectors or interlaced eigenmode projectors.

\begin{figure}
\centering
\scalebox{0.38}{\includegraphics*{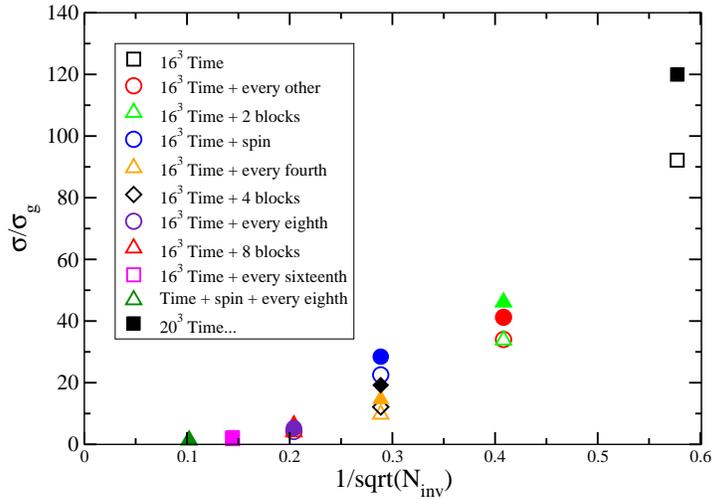}}
\caption{The statistical errors on a triply-displaced nucleon correlator at a source-sink 
separation of $5 a_{t}$ on two different lattice volumes.  The correlator was 
computed using the stochastic smearing scheme with varying levels of dilution. 
The open symbols are results obtained on a $16^{3}$ spatial lattice volume, where the 
quark-smearing operator used $32$ Laplacian eigenvectors. Closed symbols 
denote results on a $20^{3}$ spatial volume using 64 Laplacian eigenmodes.
The growth in statistical error 
which comes from increasing the spatial volume 
while the smearing cutoff and the number of quark-matrix inversions per configuration are kept fixed is modest~\cite{jbthesis}.}
\label{Fig:volcomparison}
\end{figure}

Fig.~\ref{Fig:volcomparison} compares the same error ratios for Stochastic LapH estimates of the nucleon 
correlator on two different spatial lattice volumes. Results 
for the $20^{3}$ spatial volume are plotted together with results from a
smaller, $16^{3}$ lattice. 
$32$ Laplacian eigenvectors are used to smear the quark field on the 
$16^{3}$ volume, such that eigenvalue cutoffs on the two lattices are 
approximately the same.
All other input parameters are kept fixed. 
Increasing the spatial lattice volume for a fixed eigenvalue cutoff 
and dilution scheme does lead to some increase in the statistical error. 
However, this increase is modest for all dilution schemes considered and 
the difference in errors between the two volumes decreases with higher levels of dilution.

\section{Disconnected and multi-hadron correlators}
While the results presented for the connected hadron correlators are promising, the true 
test of the Stochastic LapH method lies in its application to disconnected diagrams 
and multi-hadron correlators. 
In this section, we compare results obtained for these systems using both the Stochastic 
and exact LapH schemes. Due to the large number of quark-matrix inversions required to 
implement the exact scheme, the comparison could only be performed on the 
smaller $16^{3} \times 128$ lattice on a limited number of configurations. 
The exact method requires additional quark-matrix inversions for each sink time slice.
However, the number of quark-matrix inversions needed for the stochastic estimate is drastically reduced by applying only 
partial time dilution to the quark lines at the sink. 
In that case, the optimal dilution schemes are expected to involve projectors which are interlaced in time.

\begin{figure}
	\centering
	\ \\
	\ \\
	\scalebox{0.38}{\includegraphics*{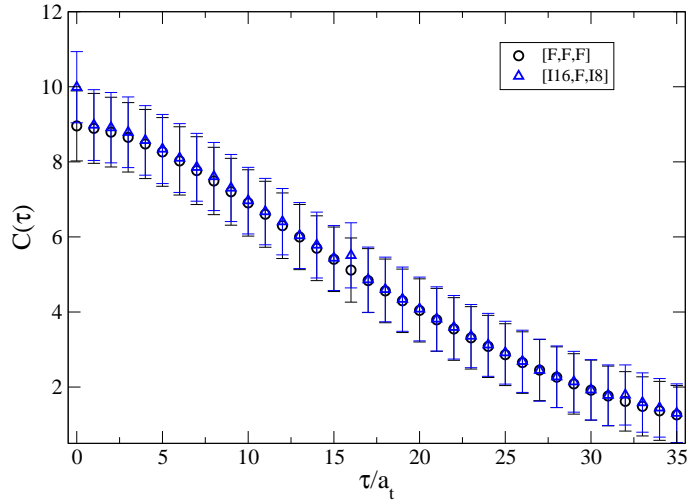}}
	\caption{Disconnected contribution to an $I=0$ pseudoscalar correlator 
	evaluated on 52 $2+1$ flavor  configurations on a $16^{3} \times 128$ lattice.
	The pion mass is approximately 400~MeV.
       	The circles are results obtained using the exact LapH smearing scheme, and the triangular
	data points are Stochastic LapH estimates. 
	The dilution scheme is interlaced $16$ in time, full spin dilution, and interlace $8$ in 
	Laplacian-eigenmode space.
        Although the error estimates from both methods are similar, for this particular 
	correlator, the stochastic approach requires a factor of $32$ fewer quark matrix 
	inversions than are needed in the exact scheme.}
       \label{eta_corr}
\end{figure}

\begin{figure}
	\centering
	\ \\
	\ \\
	\scalebox{0.38}{\includegraphics{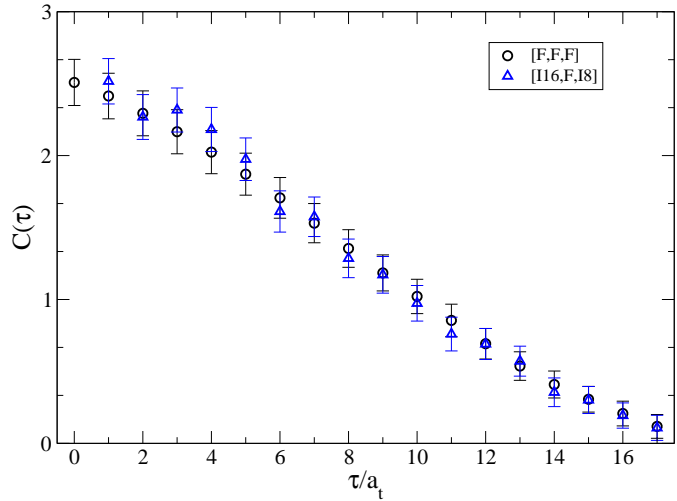}}
	\caption{Disconnected contribution to an isosinglet scalar 
	correlation function estimated using the exact and Stochastic 
	LapH methods. The correlator shown 
	was obtained after the subtraction of a large vev term.
	The discrepancy at $\tau = 0$ arises because in the Stochastic LapH estimate a single 
	diluted noise vector has been used for both the source and sink operators.
	}
	\label{sigma_corr}
\end{figure}

Disregarding contact terms, the disconnected contributions to flavor-singlet 
meson correlators can be evaluated using a single diluted noise vector.
We have computed the disconnected contributions to the isosinglet pseudoscalar and 
scalar correlators using the interpolating operators $\bar{\psi}_{l} \gamma_{5} \psi_{l}$ and
$\bar{\psi}_{l} \psi_{l}$, respectively, where the subscript $l$ denotes a light quark flavor.
Fig.~\ref{eta_corr} and Fig.~\ref{sigma_corr} show results for the disconnected contributions. 
These results were obtained on 52 configurations and the correlators were 
averaged over 128 source time slices. 
The legends list the dilution schemes used in the quark-line estimates.
[F,F,F] denotes full dilution in time, spin and eigenmode space, which corresponds to 
the exact smearing scheme. [I16,F,I8] indicates that dilution projectors are interlaced 
in time, with projectors containing every sixteenth time slice; the Stochastic LapH estimates employ full spin dilution, which was found to significantly reduce the 
variance in certain disconnected diagrams,
and interlace eight eigenmode dilution. 
In both channels, the exact smearing scheme result and the dilute Stochastic LapH estimate have 
similar errors. However, the total number of quark-matrix inversions required for the exact 
estimate is $128 \times 4 \times 32 = 16384$, while the stochastic method involves just
$16 \times 4 \times 8 = 512$ matrix inversions.

As an aside, the reader may have noticed a discrepancy between the exact and Stochastic LapH 
results for the contact term in Fig.~\ref{eta_corr}, and, more obviously, 
in Fig.~\ref{sigma_corr}. 
This is due to the use of a single noise vector for both source and 
sink operators, giving a biased correlator estimate when the temporal separation 
$\tau = 0$, but having no effect on the measurement of spectral quantities.

Note that the scalar correlator shown in Fig.~\ref{sigma_corr} was obtained after the subtraction of a large 
vacuum expectation value contribution. 
The quality of this signal, obtained on just 52 configurations, suggests that 
considerable progress can be made in the isosinglet meson sector using the Stochastic LapH
method.

In the evaluation of disconnected diagrams, it is also possible to 
use different dilution schemes for the source and sink operators. 
One can, for example, use fully time-diluted noise vectors, borrowed 
from the calculation of the connected correlator component, to 
estimate the contracted source operator in the disconnected contribution.
However, in this case, an average over all time slices is impractical. 
We tested this alternate estimate of the disconnected contribution, 
averaging the correlator over four randomly chosen source times.
However, on a moderate number of configurations using a lower level of 
time dilution but averaging correlator estimates over  all lattice time slices 
was found to give a cleaner signal, although the discrepancy between 
the estimates decreases on larger ensembles.

The inclusion of multi-hadron operators is a particular challenge for
lattice hadron spectroscopy. Not only are the quark-line diagrams much more 
complicated, but we also need to evaluate correlators involving states 
whose constitutent hadrons have non-zero momenta.

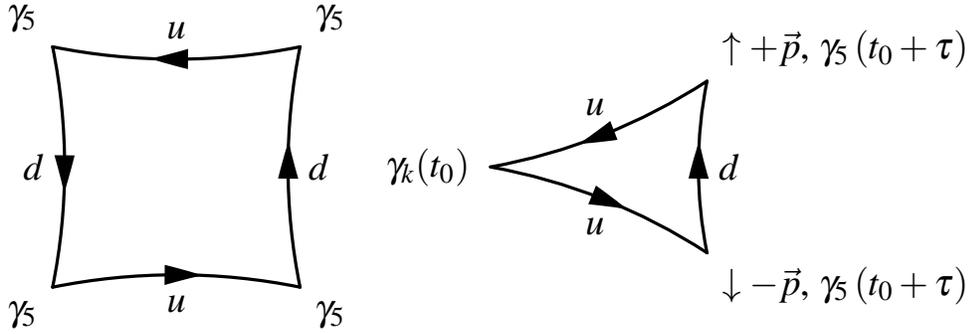
\begin{figure}
	\centering
	\begin{minipage}{2 in}
		\centering
		\ \\
		\ \\
		\scalebox{1.3}{
		\begin{fmffile}{pipi_pipi_pos}
		\begin{fmfgraph*}(90,70)
			\fmfleft{i1,i2}
			\fmflabel{$\pi$}{i1}
			\fmflabel{$\gamma_5$}{i2}
			\fmfright{o1,o2}
			\fmflabel{$\gamma_5$}{o1}
			\fmflabel{$\gamma_5$}{o2}
			\fmf{fermion, left=0.1,tension=0.2, label=$u$, label.side=right}{i1,o1} 
			\fmf{fermion, left=0.1,tension=0.2, label=$d$, label.side=right}{o1,o2} 
			\fmf{fermion, left=0.1,tension=0.2, label=$u$, label.side=right}{o2,i2}
			\fmf{fermion, left=0.1,tension=0.2, label=$d$, label.side=right}{i2,i1}
		\end{fmfgraph*}
		\end{fmffile}
		}
		\\
		\ \\
	\end{minipage}
	\hspace{0.5cm}
	\begin{minipage}{2 in}
		\centering
		\ \\
		\ \\
		\scalebox{1.3}{
		\begin{fmffile}{rho_pipi_pos}
		\begin{fmfgraph*}(70,50)
			\fmfleft{i1}
			\fmflabel{$\gamma_{k}(t_0)$}{i1}
			\fmfright{o1,o2}
			\fmflabel{$\uparrow +\vec{p}$, $\gamma_5 \left(t_0+\tau\right)$}{o2}
			\fmflabel{$\downarrow -\vec{p}$, $\gamma_5 \left(t_0+ \tau \right)$}{o1}
			\fmf{fermion, left=0.1,tension=0.2, label=$u$, label.side=right}{i1,o1} 
			\fmf{fermion, left=0.1,tension=0.2,  label=$d$, label.side=right}{o1,o2} 
			\fmf{fermion, left=0.1,tension=0.2,  label=$u$, label.side=right}{o2,i1} 
		\end{fmfgraph*}
		\end{fmffile}
		}
		\\
		\ \\
	\end{minipage}
	\caption{The two-pion box diagram and the rho-two-pion diagram discussed in the text.}\label{multi_feyn}
\end{figure}

\begin{figure}
	\centering
	\ \\
	\ \\
	\scalebox{0.38}{\includegraphics*{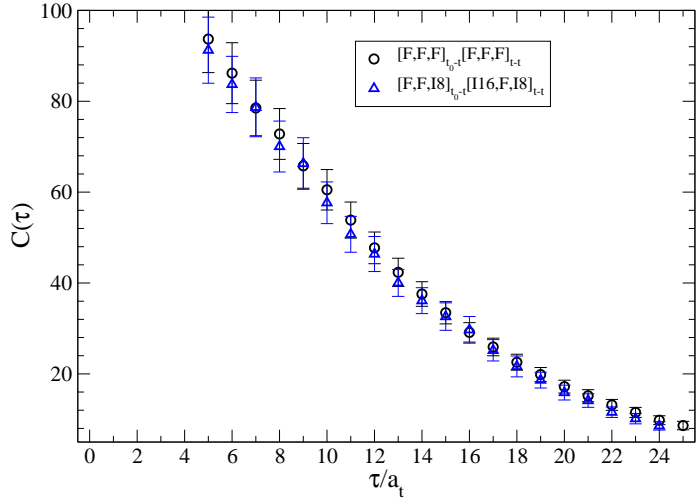}}
	\caption{The box-diagram contribution to the S-wave, two-pion correlator evaluated 
		 with exact and Stochastic LapH.
		 These results were again obtained on the $16^{3} \times 128$ lattice.
		 The correlator has been averaged over 4 well-separated source time slices.
		The stochastic estimate uses just one diluted noise vector per quark line.
	In the stochastic scheme, quark lines connecting source and sink time slices 
		are fully time dilute, while quark lines which begin and end on a single time slice 
		use the interlaced 8 time dilution scheme.
	Once again, the results from the stochastic method are as accurate as results from the exact 
	smearing scheme.
	\label{pipi_corr}
}
\end{figure}

The box diagram shown in Fig.~\ref{multi_feyn} contributes to two-pion 
correlators in the $I=0,1$ sectors.
The contribution of this diagram to an S-wave two-pion correlator 
measured on the $16^{3} \times 128$ lattice is displayed in Fig.~\ref{pipi_corr}.
The interpolating operator used is simply the product of two zero-momentum single-pion 
interpolators.
Once again, the circles denote results obtained using the exact smearing 
scheme, and the triangular data points are results from Stochastic LapH. 
In the stochastic scheme, the quark lines connecting operator 
source and sink time slices are fully time diluted, while the 
quark lines which begin and end on a single time slice employ the 
interlaced 16 time dilution scheme.
The stochastic estimate uses just one diluted noise vector per quark line.
In both cases, the correlator estimates have been averaged over four source times.

In truth, Fig.~\ref{pipi_corr} is not a fair comparison of the exact and 
Stochastic LapH schemes, since if it is feasible to apply the exact method, 
one can place source operators on all lattice time slices.
However, on larger lattices the exact smearing scheme is prohibitively 
expensive, and Stochastic LapH is the only practical means of evaluating 
multi-hadron correlation functions. 
This figure does demonstrate the potential for obtaining accurate 
estimates for multi-hadron correlation functions using Stochastic LapH.
A comparison of the error bars 
on the correlator estimates indicates that the variance in the Stochastic LapH 
estimate is dominated by gauge-field fluctuations, and no increase in 
the number of noise vectors or level of dilution is required.

As a final test of the Stochastic LapH method, we consider the two-point 
function correlating a $\rho$-meson operator at the source with and 
two-pion state at the sink. The $I=1$ two-pion system is in a P-wave 
state, which requires that the pions have non-zero back-to-back momenta.
For the source operator, we use the simple quark bilinear: $\bar{\psi} \gamma_{k} \psi$. 
This operator is correlated with an interpolating operator for two pions with one unit of momentum each. 
As in the previous examples, the stochastic estimate uses a single 
noise vector per quark line.
In this case, the Stochastic LapH algorithm gives a noticeably 
noisier result than exact smearing. However, the increase in 
error is moderate, and the stochastic result could be improved 
at negligible extra cost 
by averaging the correlator over the 
two possible ways of assigning noise vectors to the quark lines 
connecting source and sink time slices.

\begin{figure}
	\centering
	\ \\
	\ \\
	\ \\
	\scalebox{0.38}{\includegraphics{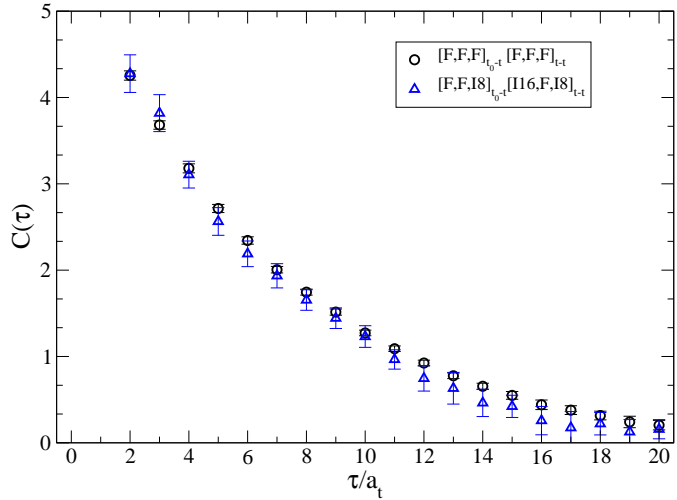}}
  \label{rho_pipi_corr}
  \caption{Two-point function correlating a $\bar{\psi} \gamma_{k} \psi$ source operator 
	  and a P-wave two-pion interpolator at the sink.
         In this case, the signal from Stochastic LapH is somewhat noisier than the exact method, although 
	still reasonable. The exact method becomes computationally intractable on larger lattice volumes.}
\end{figure}

\section{Recent developments}
Motivated by the results from the $16^{3}$ lattice, we 
 have recently begun a systematic study  of the light 
hadron spectrum on a $24^{3} \times 128$ lattice~\footnote{This corresponds to a spatial volume of approximately $(2.9~{\rm{fm}})^{3}$.}, using, in addition 
to single-hadron operators, interpolating operators for two-meson and 
meson-baryon states. To construct the required two-hadron correlators,  
we need, for each valence quark mass, independent estimates for two quark lines that begin and end on a single time slice together with 
estimates for five quark lines connecting source and sink times. 
We use a single (diluted) noise vector per quark-line estimate, and employ the same dilution schemes that were applied
to the disconnected and two-pion correlators on the $16^{3}$ lattice. 
Connected correlators have been averaged over four randomized source times. 
The number of quark matrix inversions needed to perform measurements using Stochastic LapH smearing 
is then a factor of $34$ times smaller than the number of inversions needed to implement the exact 
scheme.

The results obtained to date are promising.
In particular, a comparison of isoscalar meson and two-pion correlators measured on the $24^{3}$ spatial volume with results from the $16^{3}$ lattice confirms that the mild 
volume dependence of the Stochastic LapH method in these systems.
A detailed study of the Stochastic LapH algorithm on the $24^{3} \times 128$ lattice will 
be presented in a forthcoming publication~\cite{forthcoming}.

\section{Summary and conclusions}
We presented a new algorithm for estimating 
hadron correlation functions involving  smeared quark fields. 
Using Laplacian-Heaviside smearing, the number of quark matrix inversions needed 
to accurately estimate certain hadron correlators can be drastically reduced.
However, the strong volume dependence of the smearing scheme means that an 
exact treatment of $SM^{-1}S$ is impractical on larger lattice volumes.
We have outlined a way of implementing LapH smearing stochastically, using noise dilution to control the variance. 
Tests involving nucleon correlators show that the Stochastic LapH approach is more efficient than 
using dilute stochastic estimates for the quark propagators.
Crucially, we have found that Stochastic LapH  exhibits a mild volume dependence 
in this sector. 
The efficacy of this approach for disconnected correlators and multi-hadron correlators 
has also been demonstrated. On a small lattice volume, it was possible to compare the 
stochastic implementation of the LapH smearing scheme with the exact method. 
We found that Stochastic LapH gives results that are close to, or, in many cases, as accurate as
the results obtained from the exact method, at a fraction of the computational cost.
Finally, we noted that recent studies using Stochastic LapH performed on a larger lattice 
confirm excellent volume scaling for disconnected and multi-hadron correlators.

This algorithm is an exciting development for our spectroscopy program.
It is now possible to include multi-hadron operators in our Monte Carlo measurements, and
we can begin to extract resonance energies and widths from the measured finite-volume spectra~\cite{mike,luscher,meissner}. 
Our results indicate that the isosinglet meson spectrum is also now accessible, although in that
sector one also has to take potential mixing with glueball states into account.
The challenges that remain to properly dilineate the low-lying hadron spectrum should not 
be underestimated, but, with continued technical and theoretical advances, considerable 
progress can be made toward this goal in the near future.



\subsection{Acknowledgements}
J.F. would like to acknowledge the hospitality of the Center for Computational Sciences and the Department of Physics at the University of Tsukuba.
This work was supported by the U.S. National Science Foundation under awards PHY-0510020, PHY-0653315, PHY-0704171 and through TeraGrid resources provided by Athena at the National Institute for Computational Sciences (NICS) under grant number TG-PHY100027 and NICS and the Texas Advanced Computing Center under TG-MCA075017. M.P. is supported by Science Foundation Ireland under research grant 07/RFP/PHYF168.
We thank our colleagues within the Hadron Spectrum Collaboration. 
Numerical calculations were performed using the Chroma software suite~\cite{chroma}.

\end{document}